\documentclass[prb,superscriptaddress,preprint,endfloats,nopacs]{revtex4}%preprint type
\newcommand {\SRO}{Sr$_3$Ru$_2$O$_7$}
\newcommand {\SROinf}{SrRuO$_3$}
\newcommand {\SROone}{Sr$_2$RuO$_4$}
\newcommand {\SROthree}{Sr$_4$Ru$_3$O$_{10}$}
\newcommand {\STO}{SrTiO$_3$}
\newcommand {\ttwog}{$t_{2\mathrm{g}}$}

\newcommand {\Bout}{$B_{\parallel c}$}
\newcommand {\Bin}{$B_{\parallel a}$}
\usepackage{graphicx}
\usepackage{bm} % bold math
\usepackage{color} 
\usepackage{pifont}
\usepackage{amsmath, amsthm, amssymb}
\usepackage{dcolumn} % Align table columns on decimal point
\begin{document}

\title{Ferromagnetic state with large magnetic moments\\ realized in epitaxially strained {\SRO} films}

\author{Ren Oshima}
\affiliation{Department of Physics, Tokyo Institute of Technology, Tokyo 152-8551, Japan}
\author{Tatsuto Hatanaka}
\affiliation{Research Center for Advanced Science and Technology, the University of Tokyo, Tokyo 153-8904, Japan}
\author{Shinichi Nishihaya}
\affiliation{Department of Physics, Tokyo Institute of Technology, Tokyo 152-8551, Japan}
\author{Takuya Nomoto}
\affiliation{Research Center for Advanced Science and Technology, the University of Tokyo, Tokyo 153-8904, Japan}
\author{Markus Kriener}
\affiliation{RIKEN Center for Emergent Matter Science (CEMS), Wako 351-0198, Japan}
\author{Takahiro C. Fujita}
\affiliation{Department of Applied Physics, the University of Tokyo, Tokyo 113-8656, Japan}
\author{Masashi Kawasaki}
\affiliation{RIKEN Center for Emergent Matter Science (CEMS), Wako 351-0198, Japan}
\affiliation{Department of Applied Physics, the University of Tokyo, Tokyo 113-8656, Japan}
\author{Ryotaro Arita}
\affiliation{Research Center for Advanced Science and Technology, the University of Tokyo, Tokyo 153-8904, Japan}
\affiliation{RIKEN Center for Emergent Matter Science (CEMS), Wako 351-0198, Japan}
\author{Masaki Uchida}
\email[Author to whom correspondence should be addressed: ]{m.uchida@phys.titech.ac.jp}
\affiliation{Department of Physics, Tokyo Institute of Technology, Tokyo 152-8551, Japan}

%%%%%%%%%%%%%%%%%%%%%%%%%
%Abstract
%%%%%%%%%%%%%%%%%%%%%%%%%
\begin{abstract}
Technical advancement of oxide molecular beam epitaxy (MBE) has opened new avenues for studying various quantum transport phenomena in correlated transition-metal oxides, as exemplified by the exotic superconductivity of {\SROone} and quantum oscillations of {\SROinf}. On the other hand, film research of another Ruddlesden-Popper strontium ruthenate {\SRO} which exhibits a unique quantum phase related to metamagnetism in bulk systems did not progress well. Here we report the fabrication of high-quality {\SRO} thin films by oxide MBE and the observation of a strain-induced ferromagnetic ground state. The change in magnetic exchange coupling evaluated by first-principles calculations indicates a systematic relation between the compression of the $c$-axis length and induced ferromagnetism. Giant epitaxial strain in high-quality films will be a key to a comprehensive understanding of the magnetism in Ruddlesden-Popper strontium ruthenates Sr$_{n+1}$Ru$_n$O$_{3n+1}$, which sensitively depends on the ratio of in-plane to out-of-plane Ru-Ru distances.
\end{abstract}
\maketitle

%%%%%%%%%%%%%%%%%%%%%%%%%
%Main_Intro
%%%%%%%%%%%%%%%%%%%%%%%%%
The family of Ruddlesden-Popper (RP) strontium ruthenates, denoted as Sr$_{n+1}$Ru$_n$O$_{3n+1}$, has gathered significant interest due to its diverse quantum phases emerging at low temperatures, including a superconducting phase in {\SROone} ($n$ = 1) \cite{SRO214_firstpaper}, a nematic phase in {\SRO} ($n$ = 2) \cite{SRO327_nematic3, SRO327_nematic}, and a ferromagnetic phase with magnetic monopoles in {\SROinf} ($n$ = $\infty$) \cite{SRO113_Fang}.
The RP ruthenates consist of SrO and RuO$_2$ layers in a $n$+1 to $n$ ratio, and each repeated unit has alternately stacked SrO and RuO$_2$ with SrO termination along the $c$-axis.
As the electrons are itinerant in the RuO$_2$ layer, the electronic state of the RP ruthenates is quasi-two-dimensional for $n$ = 1 \cite{SRO214_dimensionality}, and it becomes more three-dimensional as $n$ increases \cite{SRO113_dimensionality}, leading to a variety of quantum phases formed depending on $n$.

Experimentally, the sample quality of ruthenate samples is crucial for studying these intrinsic properties of Sr$_{n+1}$Ru$_n$O$_{3n+1}$ in detail, as impurities make the observation of quantum phases more challenging \cite{SRO214_disorder, SRO327_res_mag}.
While most of the research on RP ruthenates has been performed with bulk single-crystals, recent technical advances in oxide molecular beam epitaxy (MBE) have enabled quantum transport measurements of high-quality RP ruthenate films for both end members $n$ = 1 ({\SROone}) \cite{SRO214_MBEuchida1, SRO214_Nair, SRO214_MBEuchida2, SRO214_MBEuchida3, SRO214_MBETEM} and $n$ = $\infty$ ({\SROinf}) \cite{SRO113_Nair, SRO113_takiguchi, SRO113_wakabayashi, SRO113_taiwan}.
This also enables to gain control over the parameters unique to films such as strain, thickness, and heterointerface.
For example, an enhancement of the upper critical field has been universally observed in {\SROone} films possibly due to confinement \cite{SRO214_MBEuchida2} and control of the easy-axis direction of the ferromagnetic ordering and its ferromagnetic transition temperature have been studied in {\SROinf} films by epitaxial strain \cite{SRO113_wakabayashi}.
On the other hand, {\SRO} thin film research has not yet reached such an advanced stage, strongly calling for high-quality film samples.

{\SRO} exhibits a paramagnetic ground state with strong magnetic fluctuations \cite{SRO327_nematic_mag}.
Especially in high-purity bulk single-crystal samples, it has been reported that the resistivity rapidly increases at the metamagnetic transition \cite{SRO327_nematic_mag, SRO327_nematic}.
Such behavior of the magnetoresistivity has been understood to originate from an electronic nematic phase, which spontaneously lowers the $C_4$ symmetry of the Fermi surface to $C_2$ \cite{SRO327_nematic1}.
While the formation of a spin-density wave has been suggested as an origin of the complex nematic ordering \cite{SRO327_neutron1}, the actual underlying mechanism remains still elusive, demanding advanced experiments based on high-quality films.
As in {\SROone} and {\SROinf}, the quantum phase in {\SRO} is potentially controllable by strain.
The magnetic and transport properties of RP systems could be sensitive to changes in the lattice constants, and therefore, it is very promising to engineer epitaxial strain in thin films.

While thin film growth of {\SRO} has been reported by several methods \cite{SRO327_PLDfilm, SRO327_MBEfilm1, SRO327_MBEfilm2}, the main difficulty of obtaining high-quality films stems from the thermodynamic stability of {\SRO}.
According to thermodynamics, {\SRO} exists between {\SROone} and {\SROinf} only in very narrow ranges of temperature, pressure, and flux \cite{SRO113_Nair,SRO_diagram1}, so that intergrowth of other RP structures is easily formed during the film growth, hindering the evaluation of the intrinsic properties of {\SRO}.
Here we report the successful growth of high-quality {\SRO} films on a {\STO} (001) substrate by a solid-source MBE technique.
Owing to an accurate adjustment of the Sr and Ru fluxes as well as the supply of high-purity ozone, we have achieved high-purity {\SRO} films with a limited number of intergrowths of other RPs.
Large in-plane epitaxial strain results in a transition from a paramagnetic ground state to a ferromagnetic ground state in {\SRO} films.

{\SRO} thin films were grown on (001) {\STO} step-terraced substrates using a solid-source MBE system \cite{RuO2_MBEuchida}.
The {\STO} substrate was heated at 800 $^\circ$C with a semiconductor-laser heating system, and mixed oxidant gas consisting of 60\% O$_3$ and 40\% O$_2$ was supplied at a pressure of $4.5\times 10^{-7}$ Torr from a Meidensya ozone generator.
4N Sr flux was supplied from a Knudsen cell and measured with a nude ionization gauge.
3N5 Ru flux was simultaneously supplied using an electron beam evaporator and measured with a quartz crystal microbalance, which is positioned to monitor and adjust the Ru flux even during the growth process.
Low-temperature transport measurements up to 9 T were performed using a Cryomagnetics cryostat system equipped with a superconducting magnet.
The longitudinal resistivity was measured with the conventional low-frequency lock-in technique.
The measurement was performed by a standard four-probe method with flowing an electric current along the in-plane $a$-axis and applying a magnetic field along the out-of-plane $c$-axis.
The magnetization of the film was measured using a superconducting quantum interference device magnetometer in a Quantum Design Magnetic Property Measurement System (MPMS).
Spin density functional theory (SDFT) was applied to film and bulk crystal structures at 2 K, using a k-point mesh of 8$\times$8$\times$8 and a cutoff energy of 800 eV.
Based on the SDFT results obtained by the Vienna \textit{Ab initio} Simulation Package (VASP) code \cite{VASP}, Wannier functions were constructed by the WANNIER90 code using an 8$\times$8$\times$8 k-point mesh, Sr $p$-orbitals, Ru $s$-/$p$-/$d$-orbitals\, and O $p$-orbitals \cite{Wannier90}.
The range for reproducing the band structure was set from --8 eV to 2 eV from the Fermi level.
To estimate the magnitude of the exchange interaction, the local force method was applied, and 12 k-mesh was used in the calculations.

%%%%%%%%%%%%%%%%%%%%%%%%%
%Main_Fig.1
%%%%%%%%%%%%%%%%%%%%%%%%%
Figure 1 summarizes x-ray diffraction data taken on a typical {\SRO} film.
As seen in the 2$\theta$-$\omega$ scan in Fig. 1(b), sharp peaks with Miller indices of (00\underline{2$l$}) {\SRO} ($l$: integer) are clearly observed up to high 2$\theta$ angles.
This indicates that a $c$-axis oriented {\SRO} film is epitaxially grown on the {\STO} substrate.
The film thickness is calculated to be 41.7 nm from Kiessig fringes in the reflectivity measurement shown in Fig. 1(c).

The in-plane epitaxial relation between the {\SRO} film and the {\STO} substrate is examined by performing x-ray reciprocal space mapping as shown in Fig. 1(d).
The {\SRO} (10\underline{15}) peak is slightly shifted from its original  position in bulk samples and its $q_{x}$ value becomes equal to the {\STO} (103) one.
This demonstrates that the in-plane lattice of the {\SRO} film perfectly matches that of the substrate by a tensile strain of 0.27\% along both the $a$- and the $b$-axes.
If the strain is partially relaxed with approaching the top surface, the film peak should be broadened with tail structure toward the bulk peak position, but no such behavior is confirmed.
From the $q_{z}$ value it is also found that the out-of-plane lattice of the film along the $c$-axis is compressively strained by 0.38\%.
The x-ray reciprocal space mapping was performed after transport measurements at low temperatures, also meaning that the film in-plane lattice is maintained to match the substrate one even at low temperatures.

%%%%%%%%%%%%%%%%%%%%%%%%%
%Main_Fig.2
%%%%%%%%%%%%%%%%%%%%%%%%%
Figure 2 shows cross-sectional images of the {\SRO} film taken by scanning transmission electron microscopy.
In this high-angle annular dark-field image, Sr and Ru atoms are observed as bright objects.
As can be seen more clearly in the magnification in Fig. 2(b), unit layers, each of which is composed of three SrO and two RuO$_2$ layers, are periodically stacked to form the {\SRO} lattice along the $c$-axis.
The {\SRO} lattice occupies most of the entire observation area, demonstrating the dominant growth of {\SRO} with high-crystallinity.
This is consistent with the observation of sharp x-ray diffraction peaks in Fig. 1(b).

In this {\SRO} film, the out-of-phase boundary, where SrO and RuO$_2$ layers are misaligned by a fraction of the unit layer dimension as often observed in {\SROone} films \cite{SRO214_PLD1, SRO214_PLD2, SRO214_PLDOPB}, is not observed.
By contrast, two other types of stacking faults are confirmed.
One is that a pair of SrO and RuO$_2$ layers are extra inserted or removed in the unit layer, resulting in the formation of so called Ruddlesden-Popper faults.
They are not stacked consecutively but only included as a single unit layer of {\SROthree} ($n=3$) or {\SROone} ($n=1$).
Please note that those are only hardly detected in the x-ray diffraction scan shown in Fig. 1(b).
The other is that SrO and RuO$_2$ layers are exchanged with each other leading to a vertical insertion of a SrO-SrO unit layer boundary, resulting in local formation of {\SROinf} (For details see supplementary Fig. S1).
Considering that Ruddlesden-Popper structures with other $n$ values are easily mixed even in the case of bulk synthesis \cite{SRO327_bulkgrowth}, the successful suppression of such intergrowth compounds indicates the accurate adjustment of fluxes during the growth, allowing us to study the intrinsic properties in the tensile-strained {\SRO} film.

Figure 3(a) shows the temperature dependence of the longitudinal resistivity $\rho_{xx}$.
The residual resistivity ratio (RRR), defined as the ratio between the resistivity value at 300 K and its extrapolation to 0 K, is calculated to be 24, which is the highest value among {\SRO} films reported so far. $\rho_{xx}$ under an out-of-plane field of {\Bout} = 9 T exhibits small but finite negative magnetoresistance with approaching low temperatures. This suggests that suppression of magnetic carrier scattering by the applied magnetic field becomes dominant at low temperatures. This is also in stark contrast to {\SRO} bulk, which rather exhibits conventional positive magnetoresistance derived from the Lorentz force on charge carriers in the magnetic field. Such positive magnetoresistance is confirmed even in case that RRR of {\SRO} bulk is only half of that taken for the present {\SRO} film \cite{SRO327_res_RRR12}.

To see more details, magnetoresistance data measured upon sweeping {\Bout} at various temperatures are shown in Fig. 3(b). Negative magnetoresistance accompanied by a so-called butterfly-shaped hysteresis is clearly confirmed at temperatures below 50 K. This indicates that a ferromagnetic ground state is stabilized in the {\SRO} film and ferromagnetic domains are switched by the magnetic field with hysteresis, different from the paramagnetic ground state realized in {\SRO} bulk. In the case of {\SRO} bulk, a sharp peak also appears at metamagnetic transition field of about 8 T, where the magnetic state changes from paramagnetic to forced ferromagnetic one \cite{SRO327_nematic1_SdH, SRO327_nematic_mag, SRO327_res_mag, SRO327_nematic2}. Even in case that RRR of {\SRO} bulk is only half of the present {\SRO} film one, a broad hump structure with positive magnetoresistance is observed below 10 K \cite{SRO327_res_RRR12}, which evolves into the sharp peak upon further decreasing the temperature \cite{SRO327_nematic_mag}. Therefore, the ferromagnetic behavior observed in the present film is understood to be an intrinsic property unique to {\SRO} thin films under large epitaxial strain.

The intrinsic change of the magnetic ground state is more directly confirmed by magnetization measurements.
As seen in the temperature sweep in Fig. 3(c), the magnetization $M$ steeply increases at about $T_{\text{C}}$ = 56 K for both field cooling and zero-field cooling processes, and it becomes zero again in the zero-field cooling measurement.
This indicates that a ferromagnetic transition, which is different from the one observed in {\SROinf} at about $T_{\text{C}}$ = 150 K, is induced in the {\SRO} film.
We note that a small hump in $M$ appearing around 150 K in Fig. 3(c) may be ascribed to local intergrowth of {\SROinf}, although its formation is suppressed as neither detected by transport nor by x-ray diffraction measurements.
As a possible origin of the induced ferromagnetic state, we note that the in-plane $a$- and $b$-axes of the {\SRO} film are more tensile-strained at 2 K, i.e., by 0.64\% or 0.80\% depending on their alignment with respect to the respective axes ($a$-/$b$-axis or $c$-axis) of the {\STO} substrate with tetragonal distortion.
Actually, the observed transition temperature is very close to the one reported for bulk {\SRO} under uniaxial pressure up to 0.5 GPa along the $c$-axis \cite{SRO327_uniaxialP1}, where the $a$- and $b$-axes are estimated to be elongated by 0.11\% at maximum.
While the in-plane tensile strain is much larger in the present case, it has been also reported that in uniaxial pressure experiment the transition temperature is almost independent of the magnitude of the applied pressure once the ferromagnetic state is induced \cite{SRO327_uniaxialP1}.

Figure 3(d) presents in-plane and out-of-plane magnetization curves measured at 2 K.
The tensile-strained {\SRO} film exhibits giant magnetic saturation moment $M_{\text{s}}$ of about 2 $\mu_\text{B}$/Ru with the easy magnetization direction pointing along the $c$-axis.
Taking into account the $4d^4$ electronic configuration of the Ru$^{4+}$ ion, $M_{\text{s}}$ = 2 $\mu_\text{B}$/Ru is reasonable assuming fully polarized unpaired spins in the {\ttwog} orbitals.

Figure 4(a) compares the magnetization curves of the film with those of bulk samples under uniaxial pressure applied along the $c$-axis direction \cite{SRO327_uniaxialP1}.
Bulk samples under uniaxial pressure show a ferromagnetic ground state with hysteresis, evidencing that ferromagnetism is induced by pressure, 
while without pressure they show a paramagnetic ground state with a metamagnetic transition \cite{SRO327_res_mag}.
The magnetization increases with increasing the applied pressure.

At low temperatures, unstrained bulk {\SRO} exhibits an in-plane rotation of the RuO$_6$ octahedra, as sketched in Fig. 4(b).
The magnitude of such an in-plane rotation is expected to decrease with the tensile strain.
Given such conditions and the known rotation angle of 7.18$^\circ$ reported for strain-free bulk samples \cite{SRO327_neutron5}, a rotation angle of 0$^\circ$ corresponds to an in-plane tensile strain of approximately 0.8\%.
Such a strain condition is close to the situation in the tensile-strained film at 2 K, suggesting that the RuO$_6$ octahedra in the film have almost no in-plane rotation at low temperatures as shown in Fig. 4(b).
On the other hand, the $c$-axis length is calculated to be compressed 1.14\% in the film at 2 K.

Based on the crystal structures of bulk samples and tensile-strained films at 2 K discussed above, we modeled the density of states (DOS) and the strain effect on the magnetic exchange coupling by calculations \cite{Liechtenstein1, Liechtenstein2, Liechtenstein3}.
Using the spin density functional theory method, we confirm that the largest contributions to the DOS originates from the Ru d-orbitals near the Fermi level with spin splitting (For details see supplementary Figs. S2 and S3).
Figures 4(c) and 4(d) present the exchange couplings $J$ for both cases calculated by the local force method.
A particularly important difference appears for the second nearest-neighbor (NN) Ru atoms along the $c$-axis.
This out-of-plane exchange coupling is positive and the largest also in previous first-principles calculations on {\SRO} \cite{SRO327_calc1, SRO327_calc2}.
The exchange coupling of those is increased by 28\% when their distance is reduced by 0.85\%, which we propose as the main factor enhancing the ferromagnetism in the tensile-strained film.
Such an enhancement of the ferromagnetic coupling accompanied by a shortened Ru-Ru distance along the $c$-axis is also consistent with the trends seen in Fig. 4(a).
In a recent report of {\SROinf}-{\STO} heterostructure \cite{SROhetero}, a ferromagnetic transition is observed in a Ru-bilayer system, while it is not observed in a Ru-monolayer one. This is also consistent with our argument that the exchange interaction between the out-of-plane Ru-Ru atoms is important in the magnetism of strontium ruthenates.

%%%%%%%%%%%%%%%%%%%%%%%%%
%Main_Conclusion
%%%%%%%%%%%%%%%%%%%%%%%%%
In summary, high-quality {\SRO} films with RRR = 24 are successfully epitaxially grown on a {\STO} substrate by solid-source MBE.
A change of the magnetic ground state to ferromagnetism with a saturation moment $M \sim 2\mu_{\mathrm{B}}$/Ru is induced by giant tensile strain.
First-principles calculations using the local force method highlight the importance of the ferromagnetic interaction between the Ru atoms along the $c$-axis, which is largely enhanced by the tensile strain.
The realization of a ferromagnetic ground state in tensile-strained {\SRO} indicates that the Ru-Ru distance along the $c$-axis is a key controlling parameter for magnetism in {\SRO}.
Our work also provides insight into the magnetism in the Ruddlesden-Popper series of Sr$_{n+1}$Ru$_n$O$_{3n+1}$.
Actually, {\SROinf}, with its continuous stacking of the RuO$_6$ octahedra along the $c$-axis, exhibits ferromagnetic ordering, while {\SROone}, without such a continuous stacking, does not.
Hence, the large epitaxial strain that is difficult to achieve in bulk samples will be the key in comprehensively tuning the magnetism in the Ruddlesden-Popper series in future works.

This work was supported by JSPS KAKENHI Grant Numbers JP22H04471, JP19H05825, JP21H01804, JP22H04501, JP22K18967, JP22K20353, JP22H04958 from MEXT, Japan, by JST FOREST Program Grant Number JPMJFR202N, and by The Asahi Glass Foundation, Japan.

%%%%%%%%%%%%%%%%%%%%%%%%%
%Reference
%%%%%%%%%%%%%%%%%%%%%%%%%

\newpage 

%Fig1
\begin{figure*}[t]
    \begin{center}
    \includegraphics[width=160mm]{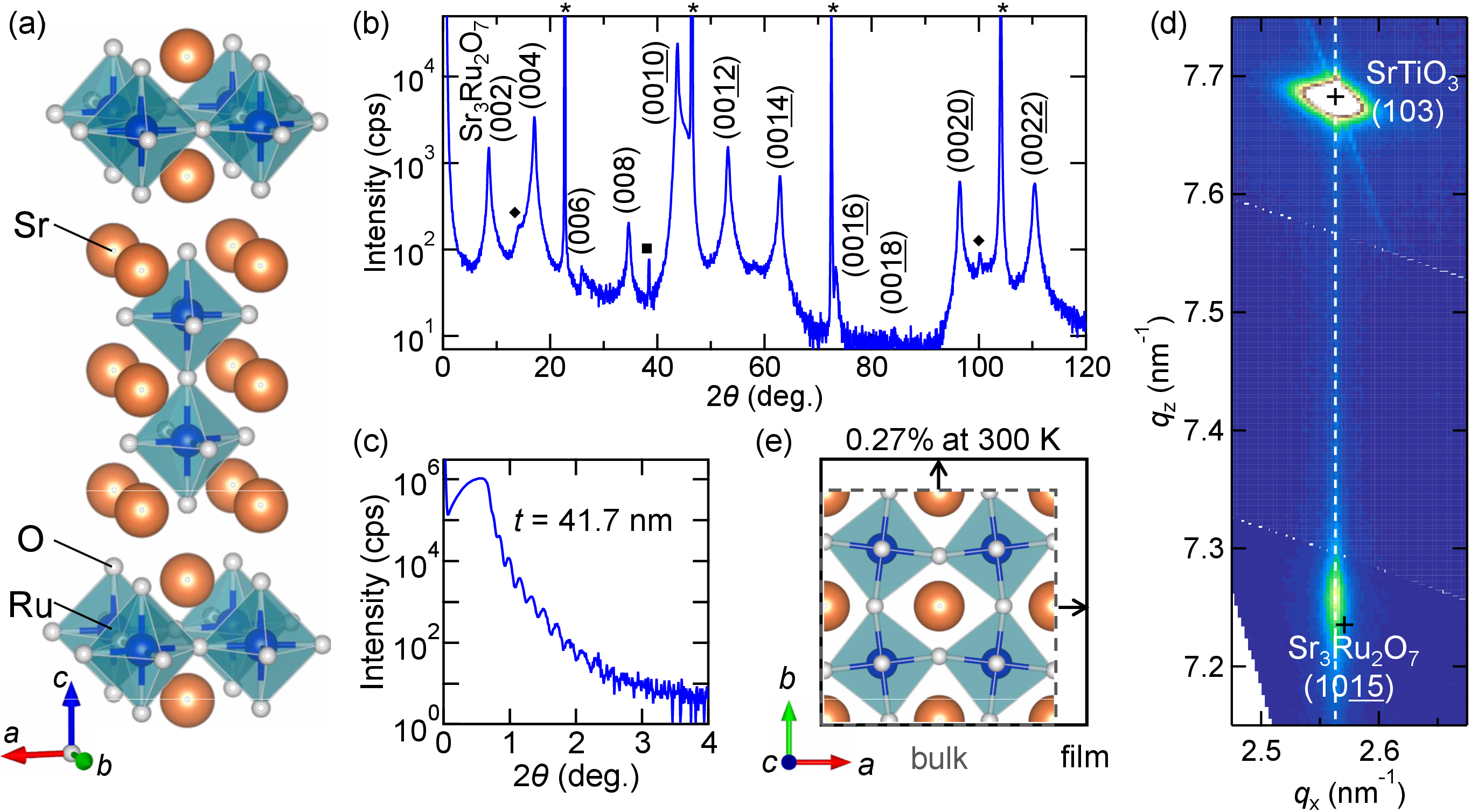}
    \caption{(color online). Epitaxial growth of {\SRO}. (a) Crystal structure of {\SRO}. (b) X-ray diffraction $2\theta$-$\omega$ scan for a {\SRO} thin film grown on (001) {\STO} substrate. Peaks marked with an asterisk, diamond, and square are assigned to {\STO} substrate, {\SROone} Ruddlesden-Popper faults, and Al metal, respectively. (c) X-ray reflectivity with clear fringes corresponding to a film thickness of 41.7 nm. (d) Reciprocal space mapping around the (10\underline{15}) {\SRO} and (103) {\STO} peaks. The in-plane lattice constant of the {\SRO} film perfectly matches that of the {\STO} substrate. (e) The in-plane lattice of bulk {\SRO} and its tensile-strained state are confirmed for the film at 300 K.}
    \label{fig1}
    \end{center}
\end{figure*}
 
%Fig2
\begin{figure}[t]
    \begin{center}
    \includegraphics[width=120mm]{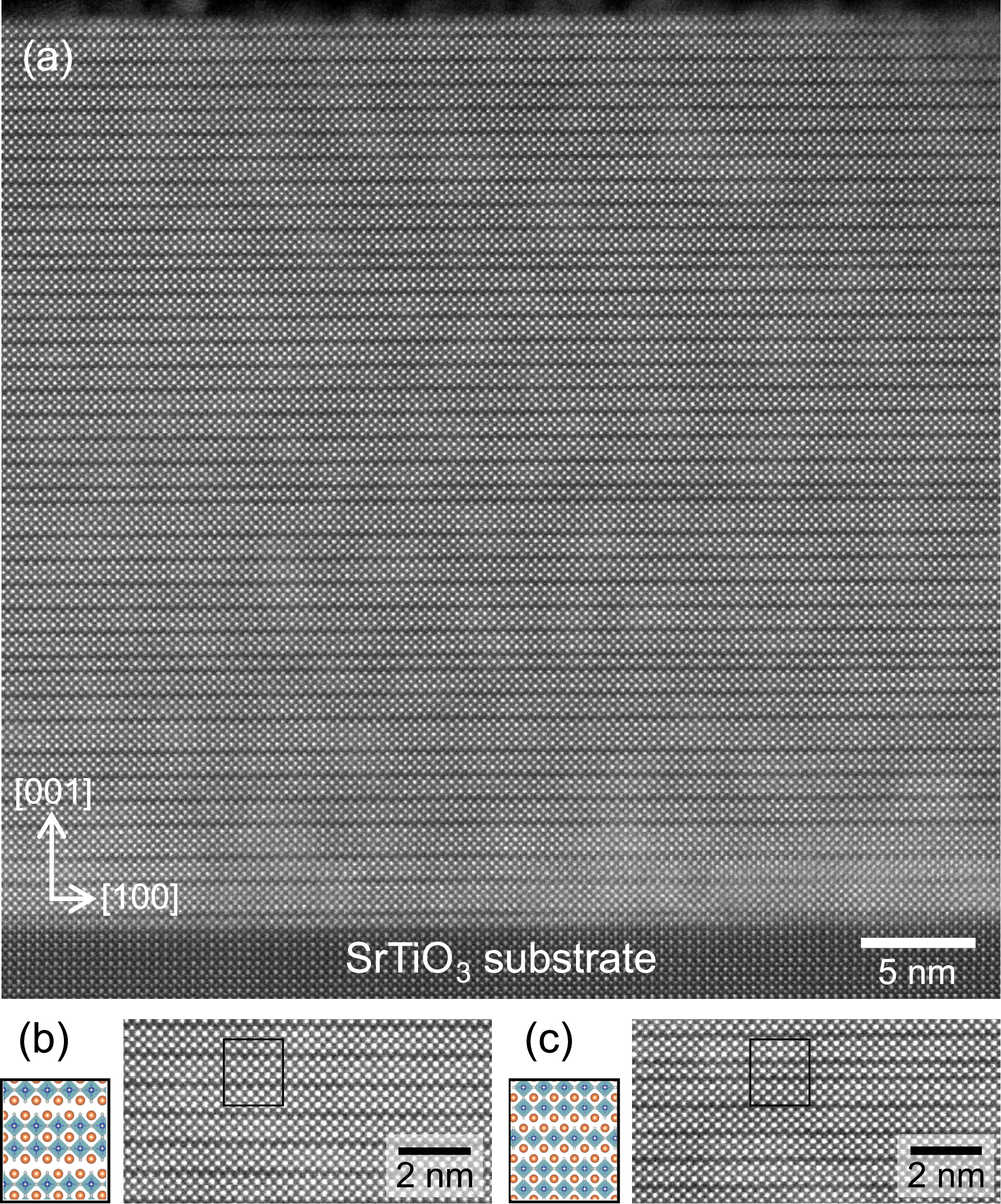}
    \caption{(color online). Atomically resolved stacking pattern. (a) Cross-sectional overview of the {\SRO} film grown on {\STO}, taken by high-angle annular dark-field scanning transmission electron microscopy. The stacking of the unit layers of {\SRO}, each composed of three SrO and two RuO$_2$ layers, is clearly resolved. Limited Ruddlesden-Popper faults are confirmed as appearance of a single {\SROthree} or {\SROone} unit layer. (b) Magnified image of the {\SRO} phase and (c) another one focusing on the insertion of a single {\SROone} unit layer.}
    \label{fig2}
    \end{center}
\end{figure}

%Fig3
\begin{figure}[t]
    \begin{center}
    \includegraphics[width=140mm]{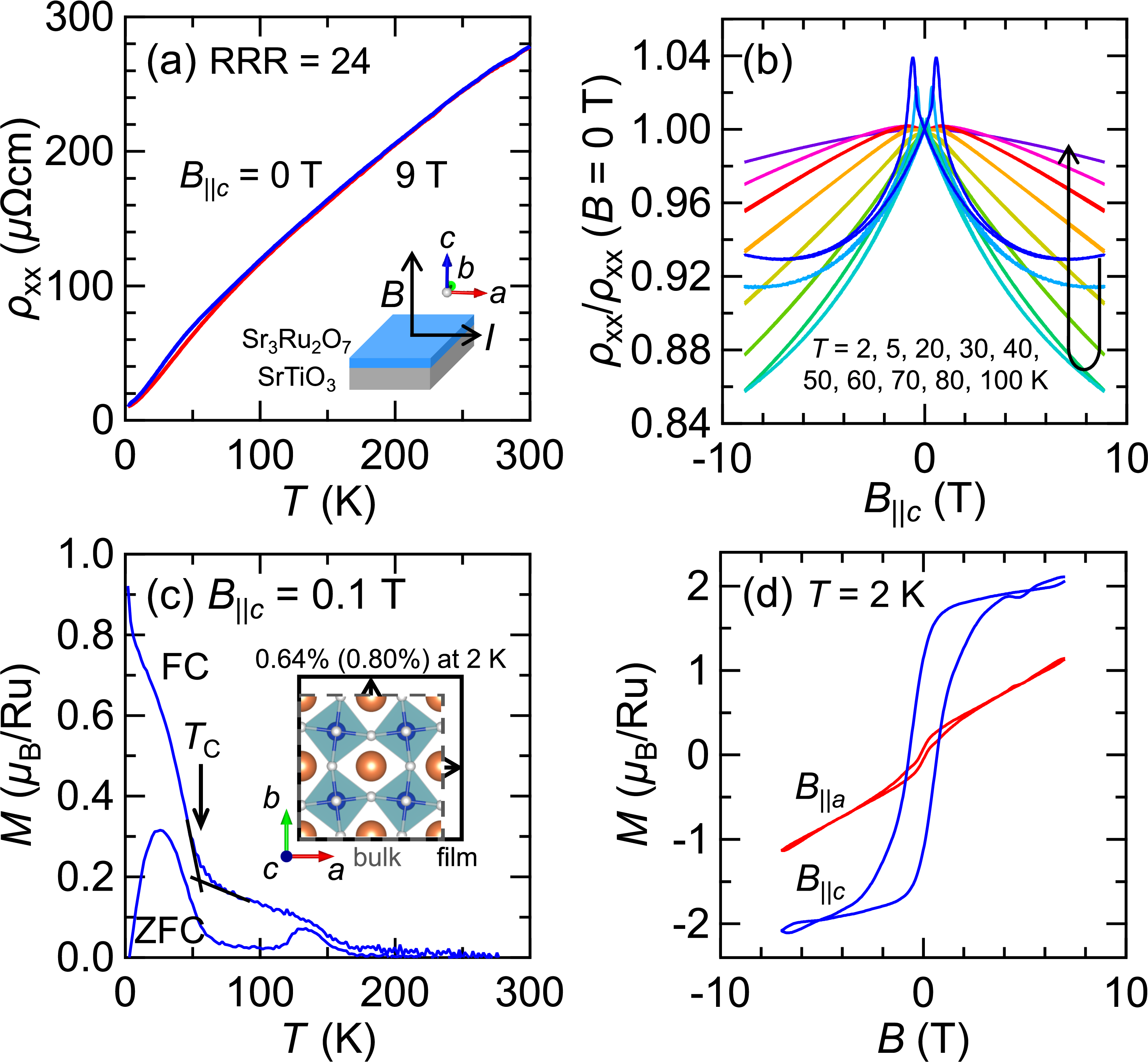}
    \caption{(color online). Ferromagnetic ground state of strained {\SRO}. (a) Longitudinal resistivity taken with applied out-of-plane magnetic fields {\Bout} = 0 T and 9 T as a function of temperature. (b) Magnetoresistance at various temperatures. (c) Temperature dependence of magnetization measured for field cooling (FC) and zero-field cooling (ZFC) processes with {\Bout} = 0.1 T, exhibiting a clear ferromagnetic transition at $T_{\text{C}} =$ 56 K. The inset depicts the $ab$-plane lattice of {\SRO} bulk and film on {\STO} at 2 K. Tensile strain from the {\STO} substrate is enhanced at low temperatures and slightly depends on the {\STO} tetragonal domains. (d) Magnetization curves for out of {\Bout} and in-plane fields {\Bin} at 2 K. A giant magnetic saturation of about $2 \mu_{\text{B}}$/Ru is confirmed for the ferromagnetic state with the easy magnetization direction along the $c$-axis.}
    \label{fig3}
    \end{center}
\end{figure}

%Fig4
\begin{figure}[t]
    \begin{center}
     \includegraphics[width=140mm]{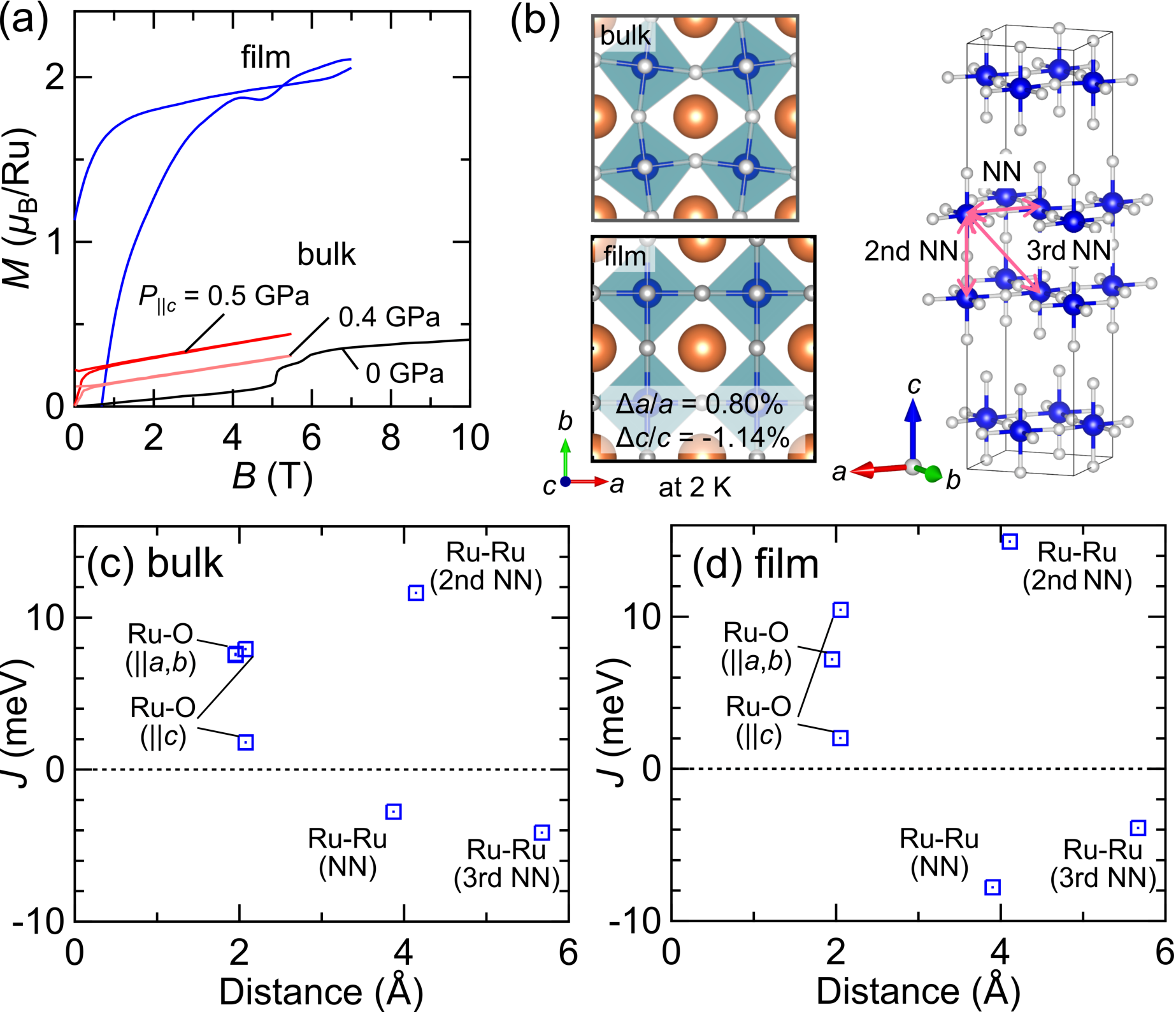}
    \caption{(color online). Theoretical calculation of the strain effect. (a) Magnetization of in-plane tensile-strained {\SRO} films and bulk {\SRO} under uniaxial pressure $P_{||c}$ of 0.0, 0.4, and 0.5 GPa \cite{SRO327_res_mag, SRO327_uniaxialP1}. (b) Crystal structures of {\SRO} bulk and film used in the calculations. Pairs of nearest-neighbor (NN), second NN, and third NN Ru atoms in {\SRO} are also highlighted. The RuO$_6$ octahedral rotation in the film is suppressed by the epitaxial strain. Exchange coupling constants calculated as a function of atomic distance for each pair of atoms in (c) bulk and (d) film.}
    \label{fig4}
    \end{center}
\end{figure}

\end{document}